\documentclass[usenatbib]{mnras}

\usepackage{newtxtext,newtxmath}

\usepackage[T1]{fontenc}
\usepackage{ae,aecompl}

\usepackage{graphicx}	
\usepackage{amsmath}	
\usepackage{amssymb}	
\title[Rate constants for the formation of SiO by radiative association]{Rate constants for the formation of SiO by radiative association}

\author[M. Cairnie, R. C. Forrey, J. F. Babb, P. C. Stancil  and B. M. McLaughlin]{
                         M. Cairnie$^{1}$\thanks{E-mail:mac6659@psu.edu},
                         R. C. Forrey$^{1}$\thanks{E-mail:rcf6@psu.edu},
                         J. F. Babb$^{2}$\thanks{E-mail: jbabb@cfa.harvard.edu},
                         P. C. Stancil$^{3}$\thanks{E-mail:stancil@physast.uga.edu}     
                          and 
                         B.  M. McLaughlin$^{2,4}$\thanks{E-mail:bmclaughlin899@btinternet.com}
\\
\\
$^{1}$Department of Physics, Penn State University, Berks Campus, 
            Reading, PA 19610-6009, USA\\
$^{2}$Institute for Theoretical Atomic, Molecular and Optical Physics (ITAMP), Harvard-Smithsonian Center for Astrophysics,\\ 
           60 Garden St., Cambridge, MA 02138, USA\\            
$^{3}$Department of Physics and Astronomy and the Center for Simulational Physics, 
            University of Georgia, Athens, Georgia 30602, USA\\
$^{4}$Centre for Theoretical Atomic Molecular and Optical Physics (CTAMOP),  School of Mathematics and Physics, \\
             Queens University Belfast, Belfast BT7 1NN, UK}

\date{Accepted XXX. Received YYY; in original form: \today}
\pubyear{2017}

\begin{document}
\label{firstpage}
\pagerange{\pageref{firstpage}--\pageref{lastpage}}
\maketitle


\begin{abstract}
{Accurate} molecular data for the low-lying states of SiO are 
computed and used to calculate rate constants
for radiative association of Si and O. Einstein A-coefficients
are also calculated for transitions between all of the bound and
quasi-bound levels for each molecular state.
The radiative widths are used together with elastic tunneling widths
to define effective radiative association rate constants which include 
both direct and indirect (inverse predissociation) formation processes. 
The indirect process is evaluated for two kinetic models which
represent limiting cases for astrophysical environments.
The first case scenario assumes an equilibrium distribution of quasi-bound states
and would be applicable whenever collisional and/or radiative excitation
mechanisms are able to maintain the population.  The second case scenario assumes 
that no excitation mechanisms are available which corresponds to 
the limit of zero radiation temperature and zero atomic density.
Rate constants for SiO formation in realistic astrophysical 
environments would presumably lie between these two limiting cases.
\end{abstract}




\section{Introduction}
The formation of dust in environments such as the 
inner winds of AGB stars \citep{Cherch2006},
the ejecta of novae \citep{RawWil89}, and 
the ejecta of supernovae \citep{Lepp1990} 
depends on the production of molecules, either 
as precursors of dust {grains} or as competitors to dust production.

The formation of CO and SiO in such environments can 
be modeled using large networks of chemical reactions, 
which depend on the knowledge of rate constants for 
individual chemical reactions relating to the molecule species.
For example,
these models have been applied in explaining the observations of 
CO \citep{Lepp1990,Gearhart1999} and SiO \citep{Liu1996} 
in the ejecta of Supernova 1987A, { the inner winds 
of AGB stars \citep{Willacy1998}, and the outer envelopes of
carbon-rich \citep{Cherchneff,Li2014} and oxygen-rich \citep{Li2016} stars.
In fact, the observation of SiO in carbon-rich 
stars suggests a shock-induced chemistry which 
dissociates CO freeing-up atomic oxygen to 
form SiO and other oxides \citep{Cherchneff}.

While in typical interstellar medium (ISM) environments, SiO
is formed primarily by the exchange reaction
\begin{equation}
\rm Si + OH \rightarrow SiO + H,
\end{equation}
the radiative association (RA) process may become
competitive, particularly in hydrogen-deficient gas.
For applications to various astrophysical environments, RA}
was studied for CO  
\citep{Dalgarno1990,Franz2011,Antipov2013}
and for SiO \citep{Andreazza1995,Forrey2016a}.

The {RA} process for SiO formation is believed 
to be a key step in the subsequent formation of 
silicates and dust \citep{Marassi2015}. Therefore, it is
of considerable interest to provide a comprehensive set
of reliable rate constants for all contributions to
the RA processes
\begin{equation}
\rm Si + O \rightarrow SiO + h\nu,
\label{direct}
\end{equation}
and
\begin{equation}
\rm Si\cdot\cdot\cdot O \rightarrow SiO + h\nu,
\label{indirect}
\end{equation}
where $h\nu$ represents an emitted photon. Here, 
the Si$\cdot\cdot\cdot$O metastable state in (\ref{indirect}) 
may be formed as an intermediate step in 
process (\ref{direct}) or by an 
independent process such as inverse predissociation.

In a recent publication \citep{Forrey2016a}, 
here referred to as paper I, our group performed detailed quantum
chemistry calculations to obtain potential energy curves (PEC's) and transition
dipole moments (TDM's) for the low-lying molecular states of SiO. 
We then used the high quality molecular data for the $X^1\Sigma^+$
and $A^1\Pi$ states, and the TDM coupling the states, 
to obtain rate constants for RA of Si($^3P$) and O($^3P$) 
via approach on the $A^1\Pi$ molecular curve. We found that the rate constants were roughly
a factor of ten smaller than those calculated earlier by \citet{Andreazza1995}.

In the present work, we extend our initial calculations
reported in Paper I to include SiO formation by RA via approach on
the $E^1\Sigma^+$, $2^1\Pi$, $3^1\Pi$, and $D^1\Delta$ molecular states.
In addition to computing the direct RA contribution (\ref{direct})
for each intermediate electronic state, 
we also calculate tunneling and radiative lifetimes 
for all of the quasi-bound vibrational states in order to provide
estimates of the indirect resonant contribution (\ref{indirect}) 
for different kinetic conditions.

In paper I, we showed that resonances may play an important role 
in enhancing the rate constants for the formation of SiO. 
With the assumption that all quasi-bound levels were in local thermodynamic
equilibrium (LTE), we obtained rate constants for low temperatures
which were several orders of magnitude larger than 
those predicted by standard quantum scattering formulations. 
The RA rate constants were defined to include both direct
and indirect (inverse predissociation) formation processes. 
The direct contribution was computed using semiclassical \citep{Bates1951} 
and  quantum mechanical methods 
\citep{Zygelman,Stancil1993,Nyman2015,Gustafsson2016,Forrey2016a} 
and excellent agreement was achieved. 
The indirect contribution relied on the kinetic LTE assumption
for the quasi-bound levels, which would be applicable whenever 
collisional and/or radiative excitation mechanisms are able 
to maintain an equilibrium population.  For low density
environments which are not subjected to substantial radiation,
the indirect contribution would clearly be less than the LTE result.

We now consider two rate constants which we believe represent 
limiting cases for RA. The first rate constant is the LTE rate constant 
described above. The second rate constant, referred to as the NLTE rate constant 
in the zero-density limit (ZDL), considers the indirect formation process for a 
non-LTE environment which has zero radiation temperature and an atomic density 
$n<<n_{cr},$ where $n_{cr}$ is the
critical density determined by equating the efficiency of RA with 
that of three-body recombination (TBR).
With this definition, the NLTE-ZDL rate constant is equivalent to
conventional methods 
that include radiative broadening in the resonance contribution when 
the tunneling width is smaller than or comparable to the radiative width
\citep{Gustafsson2016,Antipov2013,Bain1972,Bennett2003,Mrugala2003}.
For astrophysical environments with varying amounts of radiation
and mass densities, the exact phenomenological RA rate constant 
would be expected to lie somewhere in-between the LTE and NLTE-ZDL 
rate constants reported here.

In addition to presenting two rate constants for each electronic
potential curve, we also present two methods for computing the
rate constants. Both methods utilize the Sturmian approach
described previously \citep{Forrey2013,Forrey2015}. The first
method calculates the rate constant using an analytic evaluation
of the thermal average with respect to the Maxwell velocity distribution.
The second method calculates the cross section and performs the
thermal average using numerical integration. This method is
similar to conventional grid-based approaches for solving the
Schr\"odinger equation, e.g. Numerov propagation \citep{Cooley1961,Johnson1977}.
Grid-based methods generally need to be performed on a fine energy mesh in order
to fully resolve the resonant contributions. 
In the present case (Method 2), the size of the Sturmian basis set 
determines the energy density of the numerical integration grid. 
The contributions from narrow resonances 
are well-resolved due to the variational nature of diagonalizing the
Hamiltonian on an ${\cal L}^2$ basis set, and the remaining positive
energy eigenstates represent the broad resonant and non-resonant 
background contribution.

For the ${ A^1\Pi} \rightarrow X^1\Sigma^+$  transition in SiO, 
the RA cross section determined from Method 2 was compared to 
the standard perturbation theory quantum approach 
\citep{Zygelman,Stancil1993,Nyman2015,Gustafsson2016}.  
Excellent agreement was obtained for the broad features. The positions
of many of the narrow resonances were also found to be in excellent agreement.
The heights of the narrow resonances are not accurately determined by
the grid-based approach due to the well-known breakdown of 
perturbation theory \citep{Bennett2003}. The Sturmian approach
determines the heights of the resonances from the choice of kinetic model
(LTE or NLTE). Good agreement was obtained for the NLTE-ZDL rate constant
computed by the two Sturmian methods, 
and the rate constant is shown to be in good agreement with a 
semiclassical calculation at high temperatures.

\begin{figure*}
\begin{center}
\includegraphics[width=8cm]{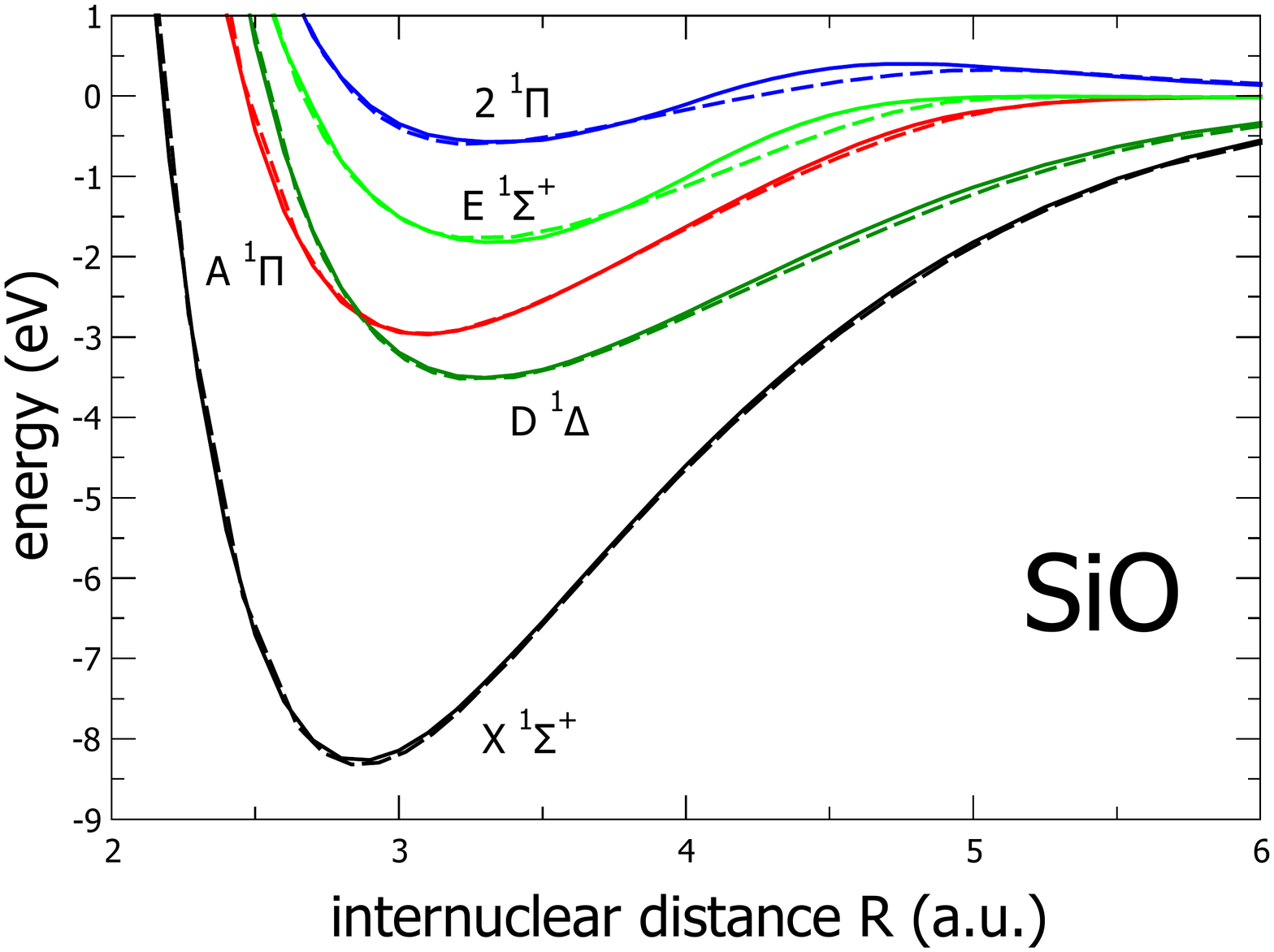}
\vspace*{.2in}
\includegraphics[width=8cm]{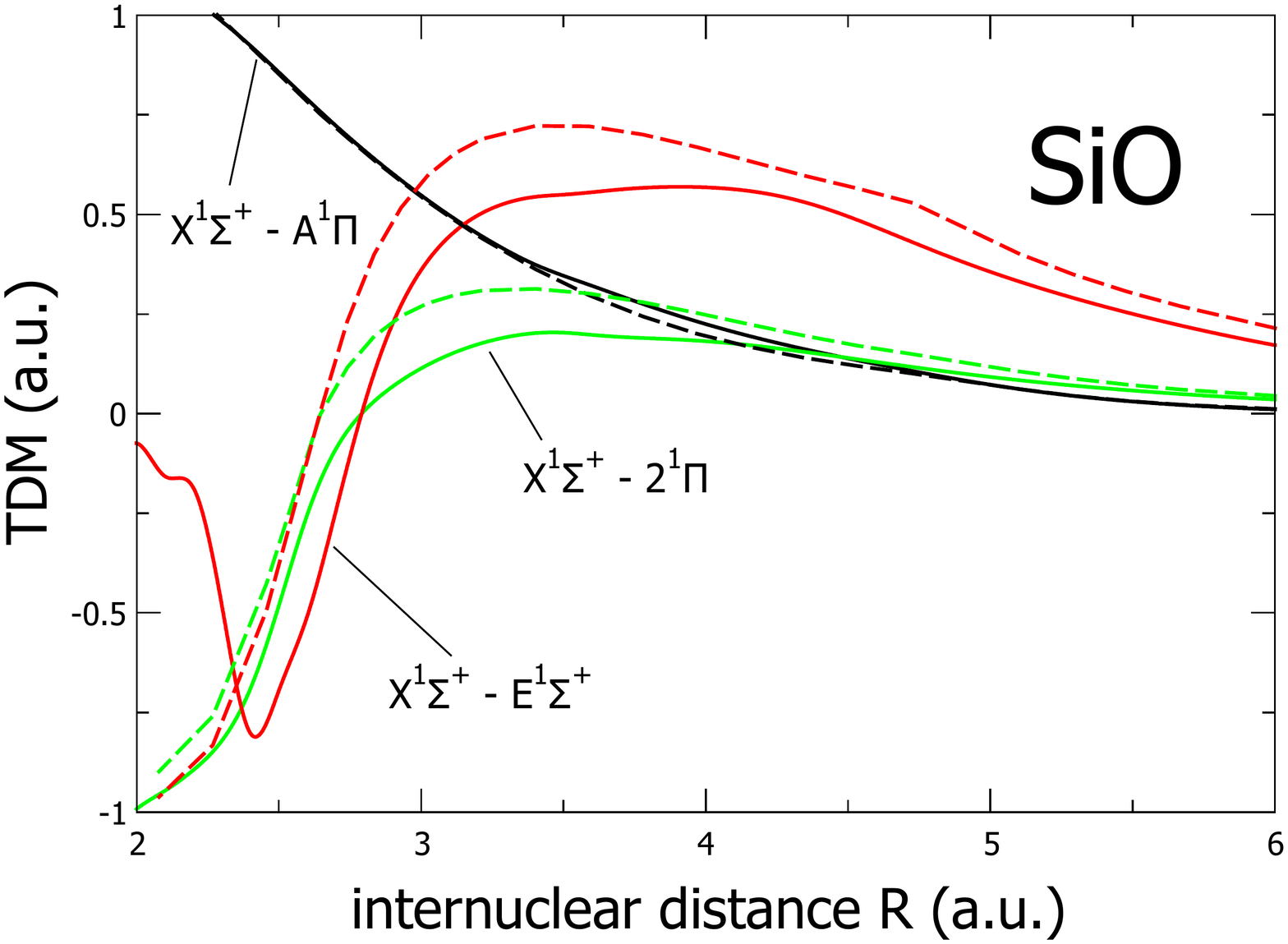}
\caption{(Colour online) 
         SiO, a comparison of the potentials energy curves (PEC's) and 
         transition dipole moments (TDM's), as a function of internuclear
         distance $R$ (a.u.), from paper I \citep{Forrey2016a} 
         (solid curves, MRCI + Q, AV6Z basis)  
         with the recent molecular electronic structure calculations of  
         \citet{CWB2016} (dashed curves, IC-MRCI, AV5Z basis).
         See text for further details. \label{fig1}}
\end{center}
\end{figure*}

\section{Theory}
\label{sec:theory}

We describe two methods for calculating the rate constant.  
Both methods utilize a Sturmian representation to form a complete 
basis set for both the dynamics and kinetics \citep{Forrey2013,Forrey2015}.
Following paper I, the RA cross section is defined by
\begin{equation}
\sigma_{\Lambda\rightarrow\Lambda^{\prime}}(E)
=\frac{\pi^2\hbar^3}{\mu E}P_{\Lambda}
\sum_{b,u}(2j_u+1)(1+\delta_u)\,
\Gamma_{u\rightarrow b}^{rad}\,\delta(E-E_u),
\label{crossx}
\end{equation}
where $\Lambda$ and $\Lambda'$ are initial and final projections of the
electronic orbital angular momentum of the molecule on the internuclear axis,
$\mu$ is the reduced mass of the Si+O system, and $E$ is the translational 
energy.  Here $b\equiv(v_b,j_b)$ and $u\equiv(v_u,j_u)$ designate
vibrational and rotational quantum numbers for bound and unbound
states, respectively. $\Gamma_{u\rightarrow b}^{rad}$ is the
probability for a radiative transition between the unit-normalized states,
and $\delta_u$ is a dimensionless parameter which may be computed
within a given kinetic model to obtain the density of unbound states.
The statistical factor
\begin{equation}
P_{\Lambda} = \frac{(2S_{mol}+1)(2-\delta_{0,\Lambda})}
{(2L_{Si} + 1)(2S_{Si} + 1)(2L_O + 1)(2S_O +1)}\ 
\end{equation}
is determined by $L_{Si}$, $S_{Si}$, $L_O$, and $S_O$, the electronic orbital
and spin angular momenta of the silicon and oxygen atoms, and $S_{mol}$
the total spin of the molecular electronic state.

\begin{figure*}
\begin{center}
\includegraphics[width=\textwidth]{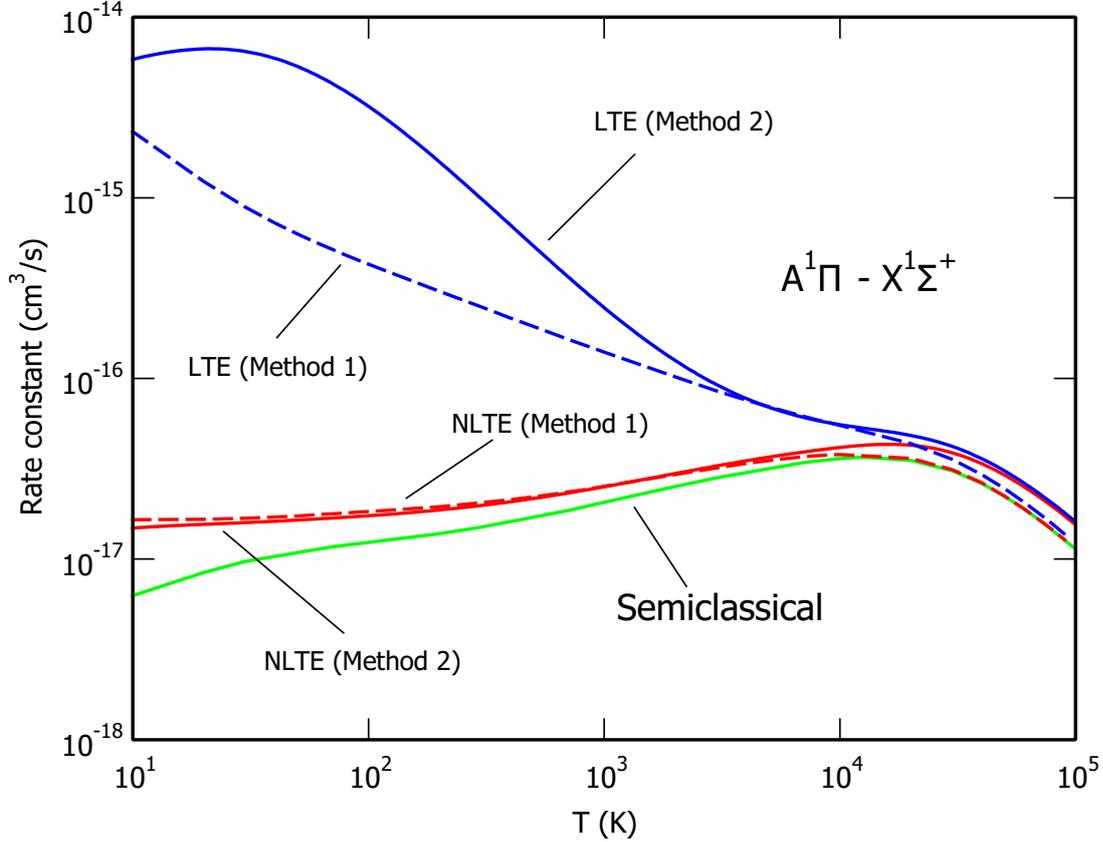}
\caption{(Colour online) Radiative association rate constants (cm$^3$/s) 
         as a function of temperature $T$ (K)
         for the $A^1\Pi\rightarrow X^1\Sigma^+$ transition in SiO.
         Method 1 refers to the analytical evaluation of the thermal average
         in equation (\ref{rateconstant}) 
         which yields equation (\ref{bob1}). 
         Method 2 refers to numerical integration of the thermal
         average in equation (\ref{rateconstant}).
         Both methods used a Sturmian basis set of 500 functions. 
         Full convergence was obtained for Method 1.
         The convergence rate for Method 2 is much slower, 
         particularly for the LTE curve which oscillates
         above the converged value. It is seen that both 
         methods converge to the semiclassical result at higher 
         temperatures, where resonance effects are no longer important. \label{fig2}}
\end{center}
\end{figure*}

The thermally averaged rate constant is defined by
\begin{equation}
k_{\Lambda\rightarrow\Lambda^{\prime}}=
\frac{1}{2\pi Q_T} \int_{0}^{\infty} E\; 
\sigma_{\Lambda\rightarrow\Lambda^{\prime}} (E)\; e^{-E/k_B T} dE\ ,
\label{rateconstant}
\end{equation}
where
\begin{equation}
Q_T=\mbox{$\left(\frac{2\pi\hbar^2}{\mu k_BT}\right)$}^{-3/2}
\end{equation}
is the translational partition function for temperature $T$
and $k_B$ is Boltzmann's constant.
Method 1 performs the thermal average analytically 
by substituting (\ref{crossx}) 
into (\ref{rateconstant}) to obtain the rate constant
\begin{equation}
k_{\Lambda\rightarrow\Lambda^{\prime}}
=\sum_{b,u}\,K_u^{eq}\,(1+\delta_u)\,
\Gamma_{u\rightarrow b}^{rad}
\label{bob1}
\end{equation}
in terms of the equilibrium constant
\begin{equation}
K_{u}^{eq}=\frac{(2j_u+1)P_{\Lambda}\exp(-E_{u}/k_BT)}{Q_T}\ .
\label{keq}
\end{equation}
This method does not require computation of the cross section.
Method 2 performs the thermal average numerically using
$\delta(E-E_u)=\delta_{E,E_u}/w_E$
to calculate the cross section. Here $w_E$ is the equivalent quadrature
weight which transforms the unit-normalized state to energy-normalization.
The weights depend on $j_u$ and may be obtained from the energy spectrum 
of the free Hamiltonian using the Heller derivative method \citep{Heller1973}. 
The Kronecker delta function reduces equation (\ref{crossx}) to a
single sum over bound states for an energy
grid comprised of unbound eigenstates of the interacting Hamiltonian.
These unbound eigenstates include quasi-bound states and discretized
continuum states, so the direct process (\ref{direct}) and the 
indirect process (\ref{indirect}) are both fully accounted for 
in the Sturmian formulation. Rate constants are computed by interpolating 
the cross section (\ref{crossx}) over the energy grid and numerically 
integrating equation (\ref{rateconstant}).

\begin{figure*}
\begin{center}
\includegraphics[width=8cm]{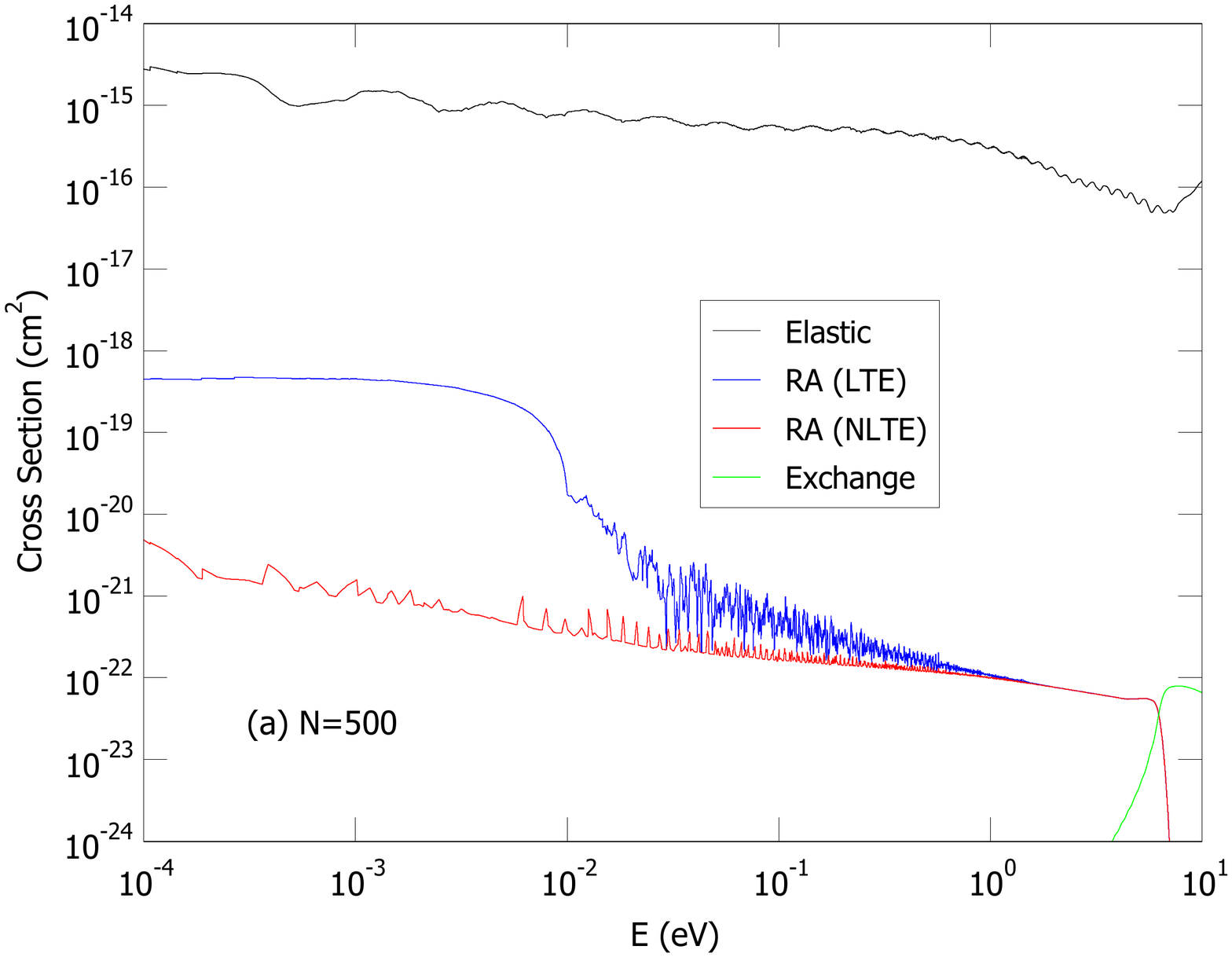}\hspace*{.1in}\includegraphics[width=8cm]{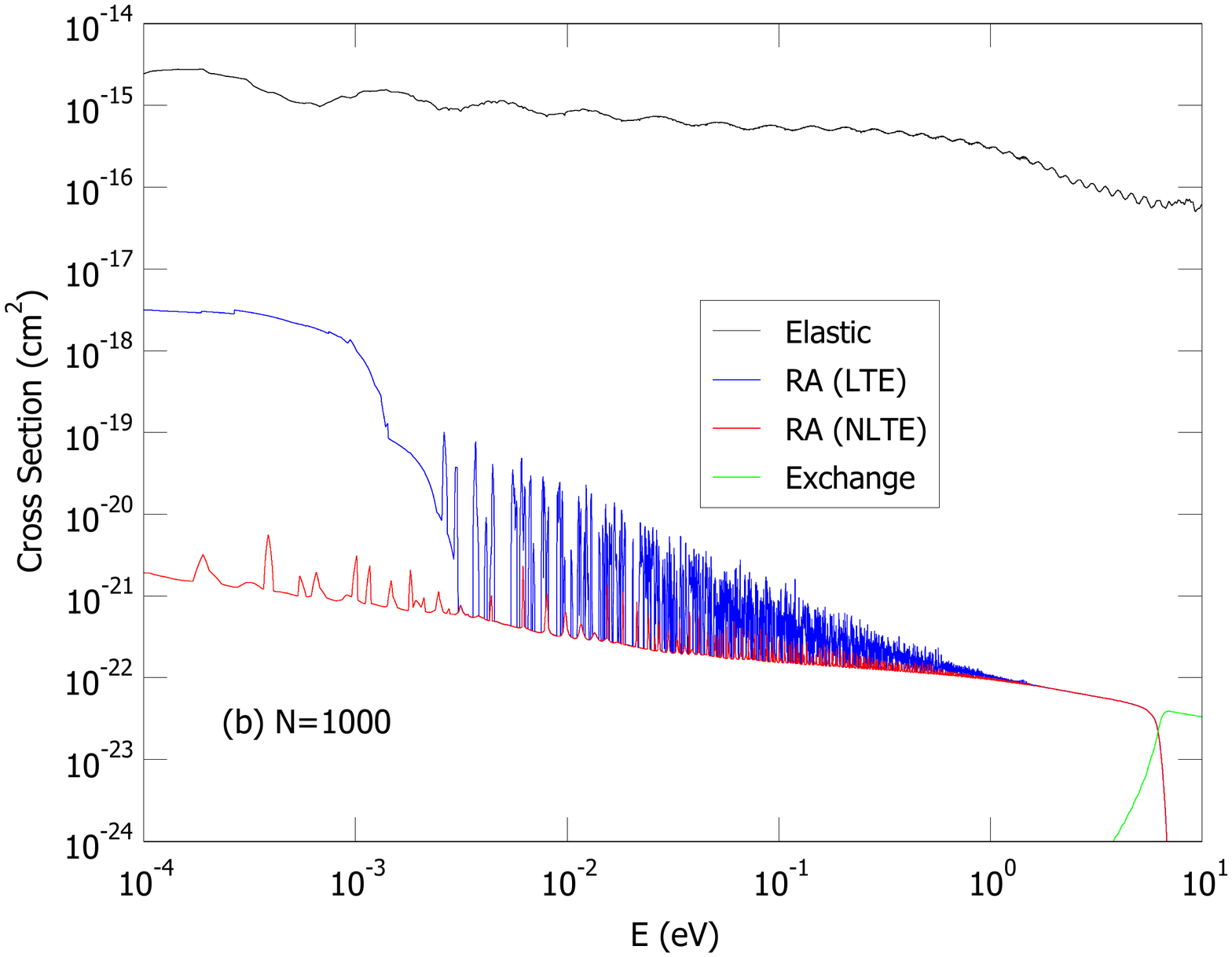}
\includegraphics[width=8cm]{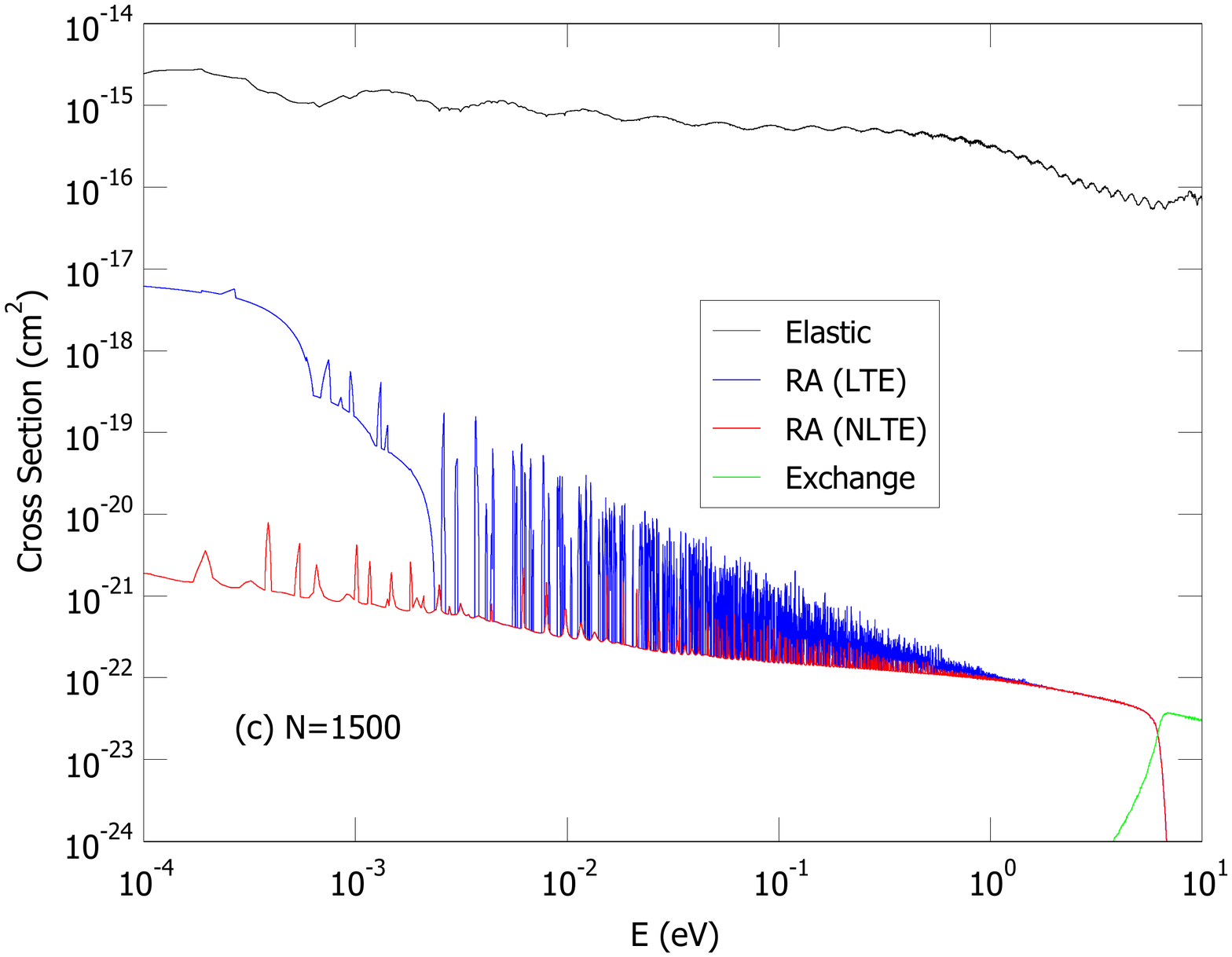}\hspace*{.1in} \includegraphics[width=8cm]{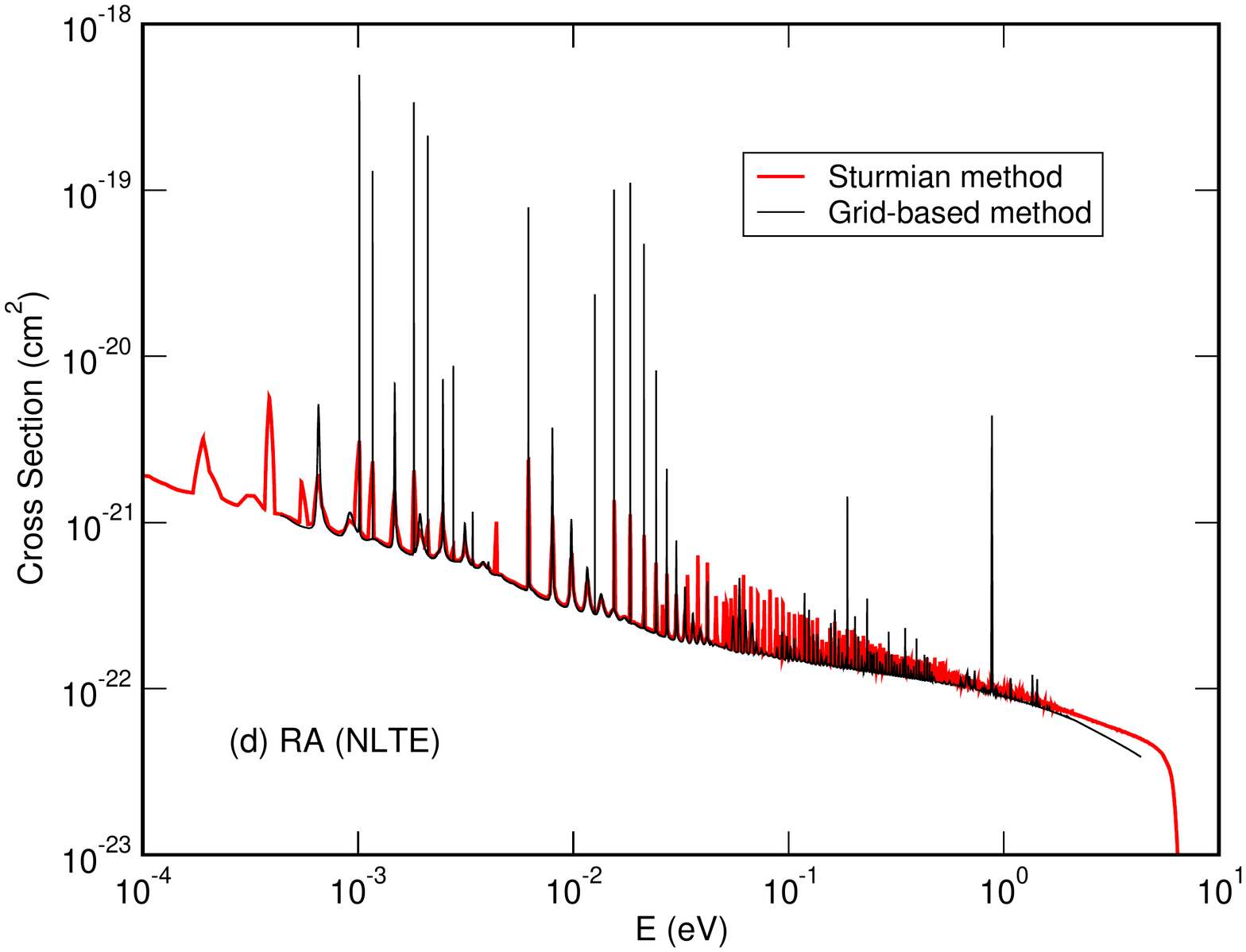}
\caption{(Colour online) 
       Cross sections for elastic scattering, RA, 
       and radiative exchange for
       approach on the $A^1\Pi$ potential. 
        The calculations used Sturmian ${\cal L}^2$ 
        basis sets with 
        (a) 500, (b) 1000, and (c) 1500 
        functions.  The resonances become better resolved 
        and more densely populated 
        as the size of the basis set is increased. 
        The RA (NLTE) cross section in (c)
        is compared against a numerical grid-based method in (d). 
        The peaks of the resonances are modulated by the 
        kinetic parameters for the Sturmian ${\cal L}^2$ 
        method but not for the quantum grid-based method.
\label{fig3} } 
\end{center}
\end{figure*}

The radiative width generally includes spontaneous and stimulated emission 
and may be written
\begin{equation}
\Gamma_{u\rightarrow b}^{rad}= \frac{A_{u\rightarrow b}}{1-e^{-(E_u-E_b)/k_BT_R}}
\label{stimulated}
\end{equation}
for a pure blackbody radiation field with temperature $T_R$.
The kinetic parameter $\delta_u$ is determined by the conditions of the gas.
For a gas in LTE, the full parameter set is defined by $\delta_{u_i}=0$.
For a NLTE-ZDL gas ($T_R=0$), the parameter set is determined by the
formula \citep{Forrey2015}
\begin{equation}
1+\delta_{u_i}=\frac{1}{1+\tau_{u_i}\left(
\sum_{j<i}A_{u_i\rightarrow u_j}+\sum_j A_{u_i\rightarrow b_j}\right)}
\label{defect}
\end{equation}
where
\begin{equation}
\tau_{u_i}^{-1}=2\pi |\langle u_i|V|f\rangle|^2
\label{width1}
\end{equation}
is the tunneling width, and $f$ is an energy-normalized free eigenstate
with the same energy as the interacting unbound state. The Einstein 
A-coefficients are given by
\begin{equation}
A_{u_i\rightarrow b_j}=\frac{4}{3c^3}(E_{u_i}-E_{b_j})^3
S_{j_{u_i}j_{b_j}}|\langle u_i|M|b_j\rangle|^2
\label{width2}
\end{equation}
and similarly for $A_{u_i\rightarrow u_j}$, where $S_{j,j^{\prime}}$ 
are the appropriate line strengths \citep{Cowan1981,Curtis2003} 
or H\"{o}nl-London factors \citep{Watson2008}, 
and $c$ is the speed of light. The electronic
dipole moment is defined by
\begin{eqnarray}
M=\left\{
\begin{array}{cc}
\langle \psi_e | p_z | \psi_e' \rangle & 
\ \ \Lambda=\Lambda'\\
\frac{1}{\sqrt{2}}\langle \psi_e | p_x+ip_y | \psi_e' \rangle & 
\ \ \Lambda\ne \Lambda'
\end{array}\right.
\label{dipole}
\end{eqnarray}
where $(p_x,p_y,p_z)$ are the components of the dipole
operator and $\psi_e$ is the electronic wave function.
We note that the $\frac{1}{\sqrt{2}}$ normalization in (\ref{dipole}) 
was double-counted in paper I for the $A^1\Pi$ to $X^1\Sigma^+$ transition.
The radiative and tunneling widths have been calculated for all of 
the {initial SiO electronic states} considered in the present work. 

\section{Results}
\label{sec:results}

The molecular electronic structure calculations used in the present work were
reported on in paper I.  In that work a multi-reference 
configuration-interaction approximation was used with the 
Davidson correction (MRCI +Q) and an aug-cc-pV6Z basis.  
The molecular orbitals were determined using the state-averaged-complete-active-space-self-consistent 
field (SA-CASSCF) approximation. Where necessary we have included updated details 
of those molecular electronic structure calculations in the present work.
Recently,  \citet{CWB2016} reported similar calculations
using the internally-contracted multi-reference {configuration-interaction} (IC-MRCI) 
approximation with an aug-cc-pV5Z basis set with 
the molecular orbitals (MO's) obtained from 
dynamically-weighted  {(DW)} CASSCF calculations.

In both paper I and \citet{CWB2016} 
the \textsc{molpro} quantum chemistry suite of molecular structure codes \citep{Werner2015} were used. 
The molecular constants for several low-lying states in SiO are compared with those obtained from the 
work of \citet{CWB2016} and previous theoretical and experimental work in Table \ref{tab1}. 
{Good agreement is found in all cases. The largest discrepancy 
occurs for the bond distances for the $A ^1\Pi$ and $E ^1\Sigma^+$, 
being $\sim$0.01 and 0.02 \AA ~larger than experiment, respectively.}

A comparison of the PEC's and TDM's 
for the two sets of calculations is illustrated in Figure \ref{fig1}. 
The solid curves correspond to the data from paper I
and the dashed curves to the data from \citet{CWB2016}. 
The PEC's are in excellent agreement for the $X^1\Sigma^+$ and $A^1\Pi$ 
states, and the agreement is also very good for the $E^1\Sigma^+$ and 
$2^1\Pi$ electronic states.  However, 
we note there is a small disagreement in the 
region around 4 -- 5 a.u. Both calculations show a
small potential barrier for the $2^1\Pi$ state. \citet{CWB2016} 
reported no barrier for this state, 
however, the data provided to us 
by the author contains a small barrier of about 0.3 eV,
{compared to a barrier height of 0.4 eV found in this work}.
The TDM's show good quantitative agreement for the 
$X^1\Sigma^+$ to $A^1\Pi$ transition, and qualitative
agreement for the $X^1\Sigma^+$ to $E^1\Sigma^+$ and
$X^1\Sigma^+$ to $2^1\Pi$ transitions.

\begin{figure}
\includegraphics[width=9cm]{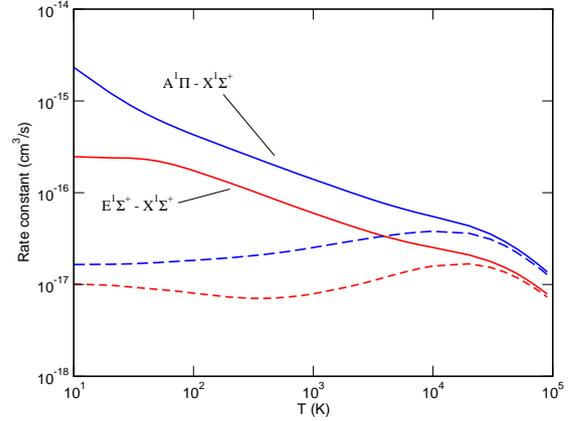}
\caption{(Colour online) Radiative association rate constants for the 
	     $E^1\Sigma^+\rightarrow X^1\Sigma^+$, and
	     $A^1\Pi\rightarrow X^1\Sigma^+$
         transitions in SiO.
	     Solid curves correspond to LTE
	     and dashed curves to NLTE. \label{fig4}}
\end{figure} 

Radiative lifetimes for the $A^1\Pi$, $E^1\Sigma^+$, and $2^1\Pi$ states 
are presented in Table \ref{lifetimes}.
These lifetimes were computed by summing equation (\ref{width2})
over all bound levels.  Excellent agreement
with previous calculations \citep{Das2003,CWB2016} is found for 
the $A^1\Pi$ state, however, the agreement is poor for the other states.
This may be due in small part to the differences in the potential curves
noted above. The larger contribution to the discrepancy 
{is attributable to the differences in} the TDM's. 
The comparison given in Figure \ref{fig1} 
for the $E^1\Sigma\rightarrow X^1\Sigma^+$ 
and $2^1\Pi\rightarrow X^1\Sigma^+$ transitions 
shows that the TDM's from paper I  
have magnitudes that are smaller than those of 
\citet{CWB2016} by amounts which are consistent with 
the discrepancies in Table \ref{lifetimes}.

\begin{table}
\centering
\caption{    Equilibrium bond distance $R_e$ (\AA{}) and
             dissociation energies $D_e$ (eV) for the $X^1\Sigma^+$, $D^1\Delta$, 
             $A^1\Pi$, $E^1\Sigma^+$, and $2^1\Pi$ states of SiO for
             the present \textsc{MRCI+Q} calculations compared to other 
             theoretical and experimental results.
             (The data are given in units conventional to
             quantum chemistry with 1~\AA{}=$10^{-10}$~m 
             and $0.529177$~\AA{} $\approx a_0$.  
             The conversion factor 
             $1.239842\times 10^{-4}$~eV = 1 
             $\mathrm{cm}^{-1}$ is also used.)
         }
\label{tab1}
\begin{tabular}{llll}
 \hline \noalign{\vskip 1mm}  
  State 	& Method  		&$R_e/$\AA{} 		& $D_e/\mathrm{eV}$\\
\hline
\noalign{\vskip 2mm}
 $X^1\Sigma^+$	&			&					& \\
		   	&MRCI+Q$^a$		&1.5153          	&8.2748\\
		  	&IC-MRCI$^b$	&1.5170		        &8.3277\\
			&MRCI+Q$^c$		&1.5100				&8.3776\\
            &MRDCI$^d$		&1.5210				&--\\
			&Experiment$^e$	&1.5097				&8.3368\\
            &Experiment$^f$	&--					&8.26 $\pm$ 0.13\\
            &Experiment$^g$ &--					&8.33 $\pm$ 0.09\\
            &Experiment$^h$ &1.5100				&8.18 $\pm$ 0.03\\
\\
  $D^1\Delta$	&			&					& \\
			&MRCI+Q$^a$    	&1.7399		        &3.4866\\
		    &IC-MRCI$^b$  	&1.7395		       	&3.5359\\
            &MRDCI$^d$		&1.7440				&--\\
            &Experiment$^e$	&1.7290				&--\\
      		&Experiment$^j$	&1.7270				&--\\
\\
 $A^1\Pi$	&				&					& \\
			&MRCI+Q$^a$    	&1.6315		      	&2.9693\\
	        &IC-MRCI$^b$  	&1.6339  			&2.9629\\
            &MRCI+Q$^c$		&1.6229    			&3.0510\\
            &MRDCI$^d$		&1.6500				&--\\
            &Experiment$^e$	&1.6206				&3.0259\\
            &Experiment$^g$	&1.6200				&2.87 $\pm$ 0.03\\
            &Experiment$^i$ &1.6199				&-- \\
            &Experiment$^j$	&1.6207				&-- \\
\\
 $E^1\Sigma^+$	&			&					& \\
			&MRCI+Q$^a$ 	&1.7625	      		&1.8177\\
			&IC-MRCI$^b$    &1.7415	       		&1.7823\\
            &MRDCI$^d$		&1.7550				&--\\
            &Experiment$^e$	&1.7398				&--\\
            &Experiment$^i$ &1.7399				&--\\
\\
 $2^1\Pi$	&				&					& \\
			&MRCI+Q$^a$ 	&1.7650	  	       	&0.5808\\
			&IC-MRCI$^b$    &1.7281	       		&0.6018\\
            &MRDCI$^d$		&1.7050				&--\\
            &Experiment$^j$	& --				&--\\
\hline
\end{tabular}
\begin{flushleft}
$^a$Multi-reference configuration interaction (MRCI) and Davidson correction (+Q), aug-cc-pV6Z basis, present work\\
$^b$Internally contracted multi-reference configuration interaction (IC-MRCI), aug-cc-pV5Z basis \citep{CWB2016}\\ 
$^c$Multi-reference configuration interaction (MRCI) and Davidson correction (+Q), aug-cc-pV6Z \citep{Shi2012}\\
$^d$MRDCI, basis, Si (7s6p5d2f/7s6p4d1f), O(4s4p1d) \citep{Das2003}\\
$^e$Experiment \citep{Huber1979}\\
$^f$Experiment \citep{Hildenbrand1972}\\
$^g$Experiment \citep{Brewer1969}\\
$^h$Experiment \citep{Gaydon1968}\\
$^i$Experiment \citep{Lager1973}\\
$^j$Experiment \citep{Field1976}\\
\end{flushleft}
\end{table}

We interpolated the \textit{ab initio} calculated PEC's and TDM's 
using cubic splines. For $R <$ 1.5 $\mathrm{a}_0$, 
the \textit{ab initio}  data was joined 
smoothly to 
the analytic form $a\exp(bR)$, where $a$ and
$b$ for each state were determined by fitting. 
For $R >$ 20 $\mathrm{a}_0$, the 
appropriate long-range forms were used for the separating atoms.
In particular, for Si($^3P$) + O($^3P$), this corresponds 
to having dispersion coefficients for the $C_5/R^{5}$ and 
$C_6/R^{6}$ terms. 
The $C_6$ dispersion coefficients were determined 
using the \citet{SK31} formula,
\begin{equation}
C_6 = \frac{3}{2} \frac{\alpha_A  \alpha_B}{ \sqrt{\alpha_A/N_A} + \sqrt{\alpha_B/N_B} }
\end{equation}
where $\alpha_A$ and $\alpha_B$ are the dipole 
polarisability with $N_A$ and $N_B$ the number of equivalent  electrons 
of the Si and O atoms, respectively. 
This yielded a value for $C_6$ of 53.67 a.u 
for the states separating to Si($^3P$) + O($^3P$), 
which is in suitable  agreement with 
the value of 63.3 a.u. determined from the  
London formula, where $C_6$ is determined via
\begin{equation}
C_6  =  \frac{3}{2} \frac{{\cal I}_A {\cal I}_B}{[{\cal I}_A + {\cal I}_B]} {\alpha_A  \alpha_B }
\end{equation}
with ${\cal I}_A$ and ${\cal I}_B$ being the first ionization energies of the 
separated atoms A and B. The $C_5$ dispersion coefficients 
were estimated using the approach 
outlined in \citet{Chang1967}, 
which respectively gave values for the $X^1\Sigma^+$ state of 25.5 a.u,
 the $A^1\Pi$ and $E^1\Sigma^+$ states are 0.0,  
 the $D^1\Delta$ state is 4.25 a.u 
 and the $2^1\Pi$ state is $-17.0$ a.u.

\begin{table}
\centering
\caption{SiO, radiative lifetimes $\tau$ (ns) for the $v=0, j=0$ level 
             of the upper electronic state. \label{lifetimes}}
\begin{tabular}{cccc}\hline
SiO ro-vibrational transition		    & $\tau$ (ns)	& $\tau$ (ns)	& $\tau$ (ns) \\
($v=0, j=0 \rightarrow v',j'$)			& CWB$^a$ 		& CCD$^b$ 		& Present work$^c$ \\ 
\hline
\ $A^1\Pi \rightarrow X^1\Sigma^+$ 		& 28.9 		& 29.3 		& 29.5 \\ 
\ $E^1\Sigma\rightarrow X^1\Sigma^+$ 	& 7.1 		& 11.0 		& 21.7 \\ 
\ $2^1\Pi\rightarrow X^1\Sigma^+$ 		& 14.9 		& 29.7 		& 90.4 \\
\hline 
\end{tabular}
\begin{flushleft}
$^a$IC-MRCI, AV5Z basis \citep{CWB2016}\\
$^b$MRDCI, (4s,4p,1d) basis \citep {Das2003}\\
$^c$MRCI + Q, AV6Z basis, present work\\
\end{flushleft}
\end{table}
In paper I, the LTE rate constants included stimulated emission
for an ideal blackbody radiation field with $T_R=T$. In the present work,
we neglect stimulated emission and assume that thermalization
is entirely due to collisions. This yields no difference between 
the two LTE definitions when $T<10^4$ K since the denominator of 
equation (\ref{stimulated}) is very close to unity for these temperatures.
The rate constants defined here, however, are smaller than 
those in paper I for $T>10^4$ K. The results presented here also correct 
the double-counting error in equation (\ref{dipole}) made in
paper I for the $A^1\Pi\rightarrow X^1\Sigma^+$ transition.
\begin{figure}
\includegraphics[width=9cm]{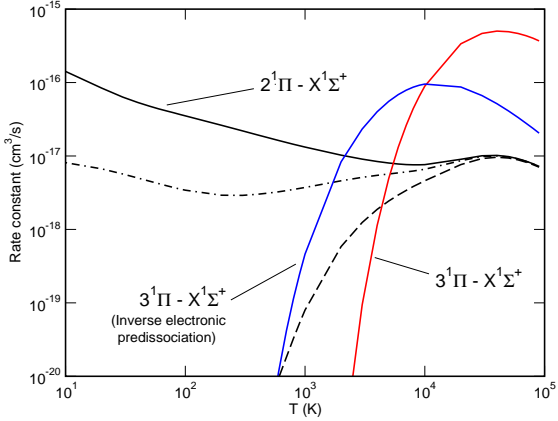}
\caption{(Colour online) 
    Radiative association rate constants for $2^1\Pi\rightarrow X^1\Sigma^+$
	and $3^1\Pi\rightarrow X^1\Sigma^+$ transitions in SiO.
	Solid curves correspond to LTE
	and broken curves to NLTE-ZDL. 
	The dashed broken curve, which uses only the
	tunneling widths for the $2^1\Pi$ potential, 
	shows a threshold due to a barrier in the potential.
	The dot-dashed curve includes tunneling widths
	from the $A^1\Pi$ potential, which removes the threshold
	for the $2^1\Pi$ rate constant.
	The rate constants for the $3^1\Pi$ state have thresholds
	that are due to the asymptotic shift in the potential.
	The red curve corresponds to approach on the $3^1\Pi$ state.
	The blue curve corresponds to a radiation-less transition
	from the $A^1\Pi$ state to an intermediate bound level 
	of the $3^1\Pi$ state prior to radiative decay to the
    $X^1\Sigma^+$ state. \label{fig5}}
\end{figure}

The RA rate constants for the $A^1\Pi\rightarrow X^1\Sigma^+$
contribution were computed using both Sturmian methods described above.
Figure \ref{fig2} shows that the NLTE-ZDL curves for the two methods are 
in reasonable agreement with each other and with the semiclassical
calculation. Method 1 is in near perfect agreement with the 
semiclassical result at high temperatures. The LTE and NLTE curves 
for both methods merge together at high temperature due to the 
diminishing importance of the resonant contribution. Discrepancies
between Method 1 and Method 2 are due to the different rates
of convergence of the two methods. Both methods used 500
Sturmian basis functions to obtain the rate constants shown 
in Figure \ref{fig2}. Full convergence was achieved for Method 1.  
The convergence rate for Method 2, however, is much slower. 

The LTE rate constant for Method 2 appears to oscillate 
above the converged value given by Method 1. This is due to 
interpolating the narrow high-$j_u$ resonances at low energies
where the energy level spacing of the Sturmian eigenvalues is 
relatively large. As a result, an artificial step-like structure 
is obtained in the cross sections at low energies. This step
goes away as the size of the basis set is increased.
Figure \ref{fig3} shows our results for basis sets 
consisting of 500, 1000, and 1500 functions.
The largest calculation for Method 2, 
which is computationally inefficient, 
has still not reached the level of 
convergence obtained by Method 1 
using far fewer basis functions. 

The LTE curve converges more slowly than the NLTE curve for 
Method 2 due to the enhanced importance of the resonances.
This is demonstrated in Figure \ref{fig3} which shows cross sections
for elastic scattering, RA, and radiative exchange (approach
on the $A^1\Pi$ curve and exit on the $X^1\Sigma^+$ curve)
which is negligible except at very high energies.
The elastic cross section is computed using
\begin{equation}
\sigma_{el}(E)
=\frac{\pi^2\hbar^2}{\mu E}P_{\Lambda}
\sum_{u}(2j_u+1)(1+\delta_u)\,
\tau_{u}^{-1}\,\delta(E-E_u).
\label{elastic}
\end{equation}
As expected, the elastic cross section is substantially larger 
than the RA cross section for all energies.
Tunneling and radiative widths are computed using equations
(\ref{width1}) and (\ref{width2}). These are used in 
equation (\ref{defect}) to obtain the set of kinetic parameters
$\delta_{u_i}$ which are needed to describe the deviation from LTE.
The kinetic parameter set is then substituted into equations (\ref{crossx})
and (\ref{elastic}) to obtain the NLTE-ZDL cross sections. 
Figure \ref{fig3} shows
that the kinetic parameters have a dramatic effect on the resonant
contribution to the cross section. Whereas the non-resonant background 
contribution converges rapidly with basis set size, 
the resonant contribution becomes better resolved and more 
densely populated as the size is increased. This is particularly
evident in the LTE curve which shows resonant enhancements that
are typically 100 times larger than the NLTE-ZDL curve. 
As shown in Figure \ref{fig2} when the cross sections are numerically 
integrated to obtain RA rate constants,
the LTE and NLTE-ZDL curves differ by about a factor of 100
at low temperatures where the resonances are most important. 
The slow convergence of the LTE rate constant illustrates 
the difficulties associated with numerically resolving 
and integrating narrow resonances.

In the Sturmian theory \citep{Forrey2013,Forrey2015}, 
the heights of the resonant peaks are determined by the 
kinetic parameters $\delta_u$. This is not the case 
for perturbation theory approaches which use a grid-based 
numerical method to solve the Schr\"odinger equation for the
continuum wavefunction.
Figure \ref{fig3}(d) compares the 
RA (NLTE-ZDL) cross section in (c) with the same 
cross section obtained by a grid-based perturbative method.
Clearly, the agreement between the two methods is excellent 
for the broad resonances and non-resonant background.
As expected, the grid-based method appears
to miss some of the narrow resonances in the dense region around
and above 0.1 eV, and the narrow resonant peaks show variable 
heights.
Both of these issues are related to the relative
spacing of the energy grid. 
A fine energy-spacing is needed to resolve the narrow resonances,
however, this can yield unphysically high
resonant peaks such as those seen in the figure.
Many of these peaks are higher than the corresponding NLTE-ZDL peaks  
and are examples of the so-called breakdown of perturbation
theory that occurs when the opacity violates unitarity
\citep{Antipov2013,Bennett2003}. The set of kinetic parameters
obtained from equation (\ref{defect}) ensures unitarity 
for a NLTE-ZDL gas
and is equivalent to conventional methods which use radiative
broadening to rescale the peaks of narrow resonances
\citep{Bain1972,Antipov2013,Bennett2003,Mrugala2003}.

Due to the faster convergence rate and convenience of bypassing
the computation of the cross section, we used Method 1 for the remaining 
calculations. The basis sets used 500 Sturmian functions with a 
scale factor of 75 a.u. All of the tunneling and radiative
widths needed to evaluate equation (\ref{defect}) were computed
and used in equation (\ref{bob1}) to obtain the RA rate constants.
Figure 4 shows the rate constants for
$E^1\Sigma^+\rightarrow X^1\Sigma^+$, and
$A^1\Pi\rightarrow X^1\Sigma^+$  transitions. The
solid curves correspond to LTE and the dashed curves to NLTE-ZDL.
The $E^1\Sigma^+\rightarrow X^1\Sigma^+$ contribution
is seen to be comparable to the $A^1\Pi\rightarrow X^1\Sigma^+$
contribution.
In both cases, the LTE and NLTE-ZDL curves are well-separated
at low temperature and gradually approach each other as the 
temperature is increased due to the diminishing importance
of the resonances.

\begin{figure}
\includegraphics[width=9cm]{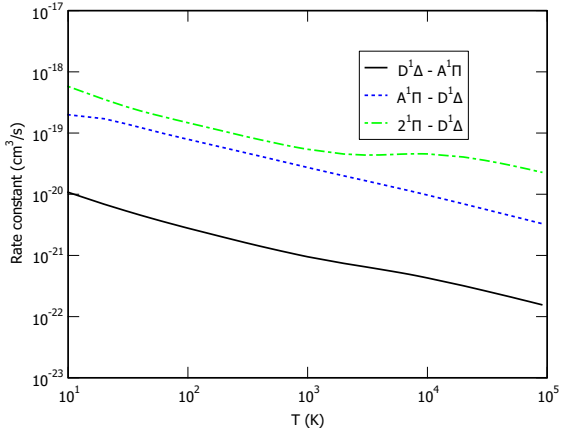}
\caption{(Colour online) 
      Radiative association rate constants for transitions in SiO.
	The dashed line (blue) is $A^1\Pi\rightarrow D^1\Delta$,
	dot dashed line (green) $2^1\Pi\rightarrow D^1\Delta$, 
	and the solid line (black) is 
     $D^1\Delta\rightarrow A^1\Pi$ 
  \label{fig6}}
\end{figure}

\begin{figure}
\includegraphics[width=9cm]{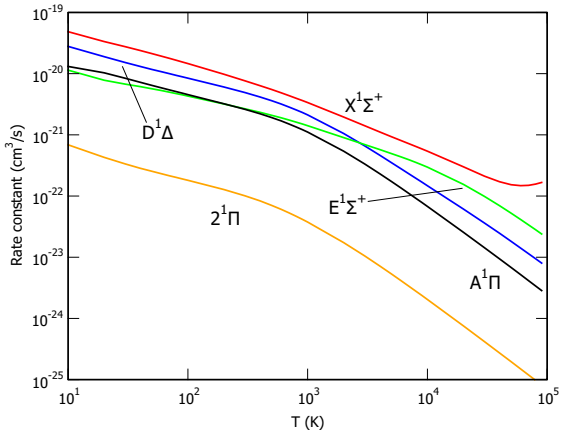}
\caption{(Colour online) Radiative association rate constants due to permanent dipoles for
	$X^1\Sigma^+$, $D^1\Delta$, $E^1\Sigma^+$, $A^1\Pi$,
	and $2^1\Pi$ in SiO. \label{fig7}}
\end{figure}

Figure \ref{fig5} shows RA rate constants 
for the $2^1\Pi\rightarrow X^1\Sigma^+$ and the 
$3^1\Pi\rightarrow X^1\Sigma^+$ transitions. Qualitatively similar 
behaviour is observed for the  $2^1\Pi$ and $A^1\Pi$ LTE curves. Two
NLTE-ZDL curves are shown for the $2^1\Pi$ state. The dashed curve
corresponds to calculations which include tunneling widths 
(\ref{width1}) for the $2^1\Pi$ potential. This curve 
shows a strong threshold effect due to a barrier in the 
potential curve \citep{Das2003,Forrey2016a}, 
which prevents low energy 
collisions from approaching close separations 
where the TDM is largest.  
The $2^1\Pi$ potential of \citet{CWB2016} 
has a {somewhat smaller barrier suggesting some
uncertainty in the rate constant given by the dashed curve.}
The dot-dashed curve corresponds to calculations which 
include tunneling widths (\ref{width1}) for both the $2^1\Pi$ 
and the $A^1\Pi$ potentials. Since the $A^1\Pi$ potential
has no barrier, the threshold is removed and the NLTE-ZDL
rate constant appears more in-line with those presented
in Figure \ref{fig4}.

\begin{figure*}
\begin{center}
\includegraphics[width=14cm]{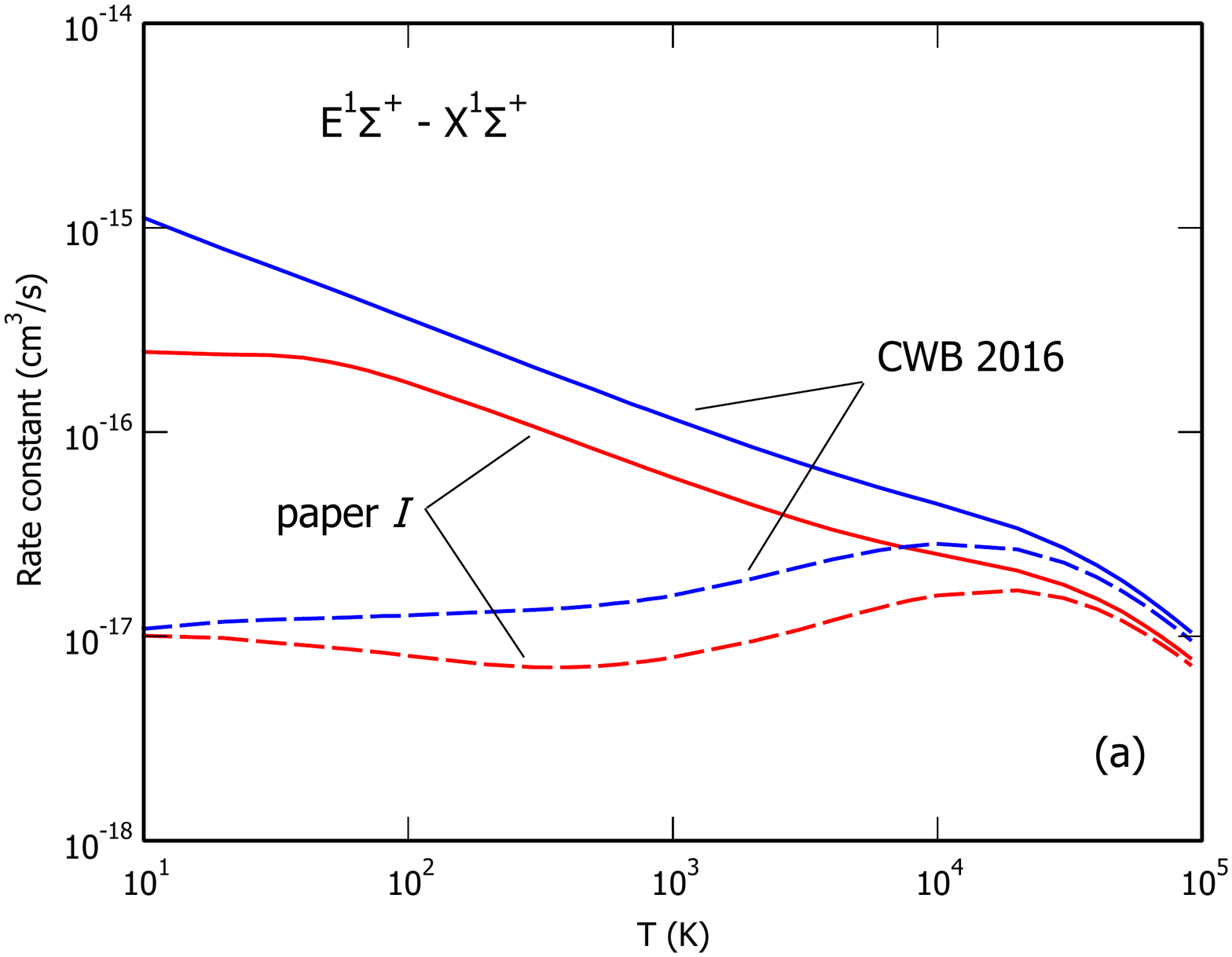}
\vspace*{.2in}
\includegraphics[width=14cm]{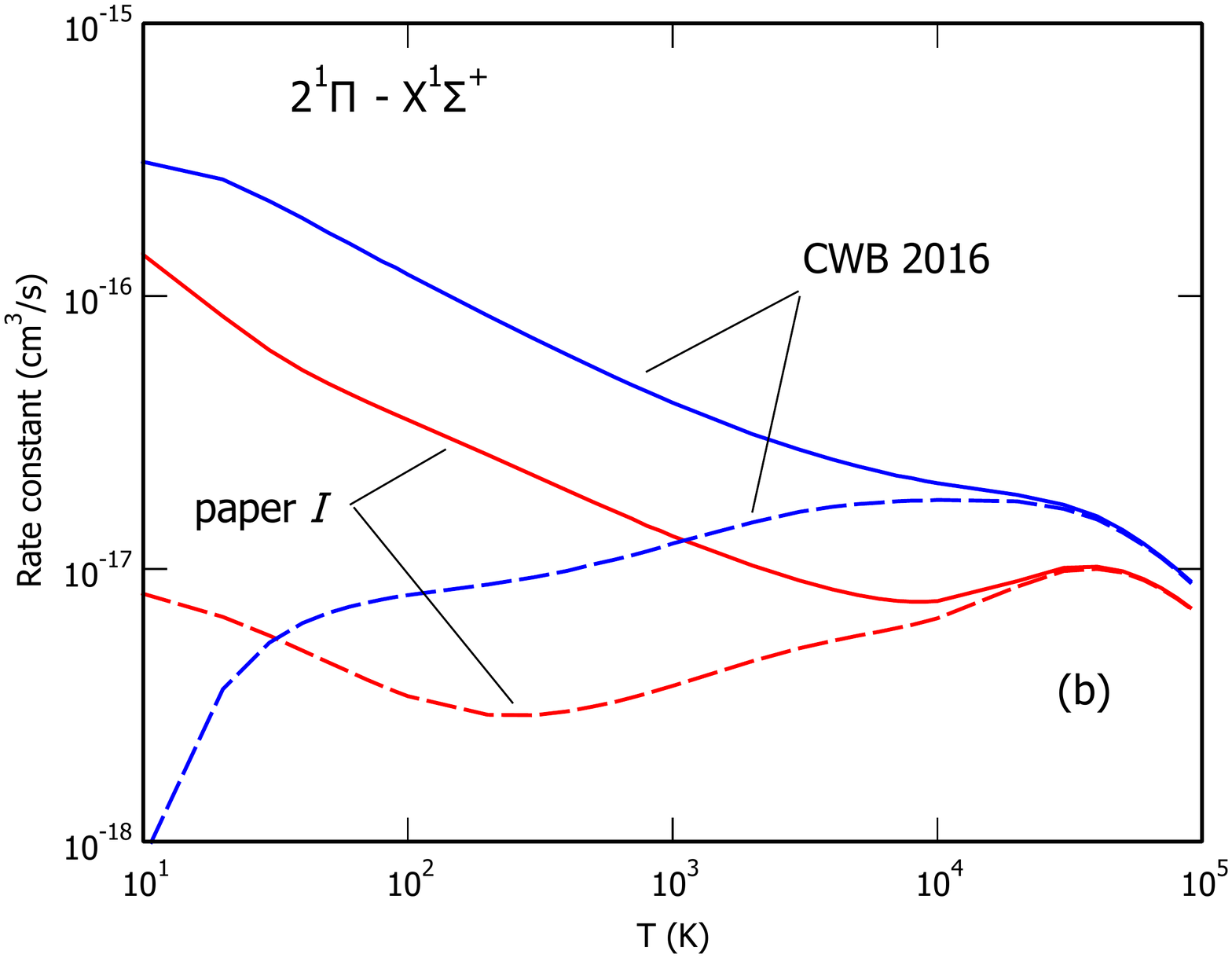}
\caption{(Colour online)A comparison of the rate constants 
         using the molecular data for PEC's and TDM's 
         from paper I  \citep{Forrey2016a}, (red curves) 
         and that from \citet{CWB2016}, (blue curves).
         (a) $E^1\Sigma^+ \rightarrow X^1\Sigma^+$ and
         (b) $2^1\Pi^+ \rightarrow X^1\Sigma^+$  transitions in SiO.
         Solid lines in both cases correspond to LTE and 
         dashed lines to NLTE-ZDL. \label{fig8}}
\end{center}
\end{figure*}

Figure \ref{fig5} also shows two RA contributions from the $3^1\Pi$ state.
The $3^1\Pi$ state separates at long internuclear distance to Si($^1D$)+O($^1D$), 
which is shifted from the Si($^3P$)+O($^3P$) asymptote by 2.7 eV. 
This produces a threshold for RA via the $3^1\Pi$ 
state (red curve) which is at a sufficiently high energy 
such that resonances do not make a substantial contribution, and that 
the LTE and NLTE-ZDL rate constants are virtually identical. 
It should be noted that the $3^1\Pi$ state also allows RA to take place 
through inverse electronic predissociation. 
In this process, bound ro-vibrational levels of the $3^1\Pi$ state 
are populated by approach on the $A^1\Pi$ state, characterized by
the tunneling widths (\ref{width1}). Subsequently, the bound levels 
of the $3^1\Pi$ state can decay to the $X^1\Sigma^+$ in a similar manner 
to the quasi-bound rotational states which contribute to RA through
inverse rotational predissociation. When the translational energy
of the approaching atoms is large enough to match the energy of
the bound levels of the $3^1\Pi$ state, the rate constant is 
non-zero and rises rapidly (blue curve). 
It is noteworthy that the bottom of the potential
well for the $3^1\Pi$ state has an energy which is approximately
the same as the energy of the potential barrier for the $2^1\Pi$ state.
This causes the threshold for the $3^1\Pi$ inverse
electronic predissociation curve to be at about
the same temperature as the threshold for the
$2^1\Pi$ NLTE-ZDL curve. The relatively large maximum
for the red $3^1\Pi$ curve is due to large energy differences 
involved in the radiative transitions,
and the replacement of $P_\Pi=2/81$ with $P_\Pi=2/25$ due to the
different atomic asymptotes. The blue $3^1\Pi$ curve does not
undergo this statistical replacement since the atoms approach
each other on the $A^1\Pi$ state.

\begin{table*}
\centering
\caption{Parameters for the analytic formula (\ref{fit}) used 
        to fit  the SiO rate coefficients as a function 
        of temperature $T$ (K) over the range 10 -- 10,000 K.\label{tab-fit}}
	    \vspace*{.2in}
\begin{tabular}{ccccccc}
\hline
Fitting &  & \hspace*{-.8in} $A^1\Pi\rightarrow X^1\Sigma^+$ \  &
& \hspace*{-.8in} $E^1\Sigma\rightarrow X^1\Sigma^+$ \  &
& \hspace*{-.8in} $2^1\Pi\rightarrow X^1\Sigma^+$ \  \\ 
\ parameter \  
& \hspace*{.0in}  LTE & \hspace*{.2in} NLTE-ZDL \hspace*{.2in}
& \hspace*{.0in}  LTE & \hspace*{.2in} NLTE-ZDL \hspace*{.2in}
& \hspace*{.0in}  LTE & \hspace*{.2in} NLTE-ZDL \hspace*{.2in} \\ 
\hline
$a$ 	& 2.08233  & 0.222128  & 0.816526    &  0.0846568   & 0.244134 & 0.0360735 \\ 
$b$ 	& 0.420491 & $-$0.161335 &  0.373286 & $-$0.0982334 & 0.488114 & $-$0.168372 \\ 
$c_1$ 	& 481.186  & 5.71969   &  879.858    &  7.91451     & $-$174.301 & 9.10961 \\ 
$c_2$ 	& 0        & 0         & $-$8.44818  & $-$1191.94   & $-$695.992 & $-$237.498 \\ 
$c_3$ 	& 0        & 0         & $-$21.8148  & 273212       & 150707  & 11728.5 \\ 
$d_1$ 	& 1.37764  & 14.09     & 78.2256     &  18.4612     & 45.4479 & 15.4816 \\ 
$d_2$ 	& 0        & 0         & 0.3277      &  830.387     & 388.013 & 536.223 \\ 
$d_3$ 	& 0        & 0         & 3.05515     &  19999.8     & 18243.8 & 12844.6 \\ 
\hline
\end{tabular} 
\end{table*}

Figure \ref{fig6} shows RA rate constants for
$A^1\Pi\rightarrow D^1\Delta$,
$2^1\Pi\rightarrow D^1\Delta$, and $D^1\Delta\rightarrow A^1\Pi$
transitions. These rate constants are smaller than those 
which lead to the formation of bound states for the $X^1\Sigma^+$ potential.
The $A^1\Pi$, $D^1\Delta$, and $E^1\Sigma^+$ potentials also support bound states, 
however, {the dissociation energies of these electronic states 
are less than that of the $X^1\Sigma^+$.} 
Consequently, the $(\Delta E)^3$ factor in equation (\ref{width2})
is not as {large} and the RA rate constants are smaller.  The three curves
shown in Figure \ref{fig6} are the largest among all transitions which
do not include $X^1\Sigma^+$ bound states. These curves represent
the LTE rate constants; the NLTE-ZDL rate constants have the usual 
fall-off behaviour with decreasing temperature and are not shown.
RA rate constants for $D^1\Delta$ to $E^1\Sigma^+$ 
and $X^1\Sigma^+$ to $A^1\Pi$, 
$D^1\Delta$, and $E^1\Sigma^+$ were computed and found to be smaller
than $10^{-25}$ cm$^3$/s.

Figure \ref{fig7} shows permanent dipole contributions for the $X^1\Sigma^+$,
$E^1\Sigma^+$, $A^1\Pi$, $2^1\Pi$, and $D^1\Delta$ states. Again, 
only the LTE rate constants are shown due to the relatively
small values of the NLTE-ZDL rate constants at low temperatures.
The LTE rate constants are comparable in magnitude to the
transition dipole contributions shown in Figure 6.

Having considered all possible RA contributions, 
it is clear that formation of SiO is dominated
by the $A^1\Pi$, $E^1\Sigma^+$, and $2^1\Pi$ states at 
low temperatures, and by the $3^1\Pi$ state at high temperatures.
In order to estimate the uncertainty in the RA rate constants
for these states, we compare results using the molecular data
of \citet{CWB2016} with those obtained using data
from paper I. As expected from the comparisons 
shown in Figure 1, the rate constants for the $A^1\Pi$ to $X^1\Sigma^+$ 
transition were virtually identical, so the comparison
is not shown. The $E^1\Sigma^+$ to $X^1\Sigma^+$ and
$2^1\Pi$ to $X^1\Sigma^+$ comparisons are shown in Figure \ref{fig8}.
Consistent with the lifetimes presented  in Table \ref{lifetimes}, the RA rate constants
calculated  with the molecular data of \citet{CWB2016} tend to be
about 2-3 times larger than those obtained using the molecular data from paper I.
It should be noted that the appropriate long-range forms for each state were 
included for the molecular curves from paper I but not for the CWB potentials.
The rate constants are found to be sensitive to the long-range for $T<100$ K
(see supplementary material). The enhanced sensitivity of the NLTE-ZDL rate constant 
in Figure 8(b) is due to the barrier in the 2$^1\Pi$ potential
and the coupling to the A$^1\Pi$ state.

\section{Summary and Conclusions}

Two methods were used to compute RA rate constants 
for $A^1\Pi$ to $X^1\Sigma^+$ transitions. Both methods
employed Sturmian basis sets to represent the bound and
unbound ro-vibrational states. Results for the two methods 
are found to be in good agreement, however, Method 1 
converges more rapidly and does not require calculation
of the cross section. This method was subsequently used
to calculate  LTE and NLTE-ZDL rate constants for the 
low-lying electronic states of SiO.  
The LTE rate constants assume
that all unbound states, including long-lived quasi-bound
states, are populated by a Boltzmann equilibrium distribution. 
The NLTE-ZDL rate constants refer to a dilute gas at $T_R=0$ which have 
no excitation mechanisms. More general NLTE environments which have
some collisional and/or radiative excitation capability would
presumably have rate coefficients which lie somewhere in-between
the NLTE-ZDL and LTE curves. 

In light of the expectation that the true formation rate constant
lies in-between the LTE and NLTE-ZDL values, it is tempting to
provide a formula which interpolates between the two limits. 
This approach has been adopted previously 
\citep{Lepp1983,Martin1996,Glover2008} 
for interpolating  collision-induced dissociation rates 
between low and high density.

However, it should 
be noted, that there are two distinct ways 
to achieve the LTE results given here. The first assumes that
the density is large enough that the quasi-bound states may be 
populated by three-body collisions. In this case, the 
total recombination rate constant
\begin{equation}
k_r=k_r^{(2)}+k_r^{(3)}n
\end{equation}
would likely be dominated by the TBR contribution $k_r^{(3)}$. 
Therefore, the LTE RA rate constant $k_r^{(2)}$ may be used to estimate 
the TBR rate at the critical density $n_{cr}$, and it is possible
that an interpolating scheme between low and high densities would
be useful. The second way to
achieve the LTE RA rate constant is to assume a pure blackbody
radiation field with $T_R=T$ \citep{Forrey2015}.
Realistic radiation fields are
generally less intense than a pure blackbody field, so a 
formula for interpolating between $T_R=0$ and $T_R=T$
would need to account for the dilution.
See \citet{Ramaker1979} for an example 
where such an approach would be useful.

A simpler approach would be to use both the LTE and NLTE-ZDL
rate constants in separate calculations to provide theoretical 
error bars in any model which depends on the SiO formation rate.
To facilitate this approach, we provide analytic fitting functions 
\citep{Novo13,Vissa16}
for both LTE and NLTE-ZDL rate constants using the form
\begin{equation}
k_r=\left[a(400/T)^b+T^{-3/2}\sum_{i=1}^3 c_i\exp(-d_i/T)\right]
\times 10^{-16}\ \mbox{cm}^3/\mbox{s}
\label{fit}
\end{equation}
where the parameters are given in Table \ref{tab-fit}. This  
formula provides a fit to the rates to better than 10 percent 
over the temperature range 10 -- 10,000 K.
A plot which compares the fitted rates with the calculated rates
is included in the supplemental material. Also included in the
supplementary data are the molecular data used to compute the
rates. The primary source of error in the calculations is due 
to the TDM uncertainties (see Fig. 1b) which yield rate
constants that differ by 2-3 times over the temperature range
considered.

\section*{Acknowledgements}
{MC and RCF} acknowledge support from NSF Grant No. PHY-1503615. 
PCS acknowledges support from NASA grant NNX16AF09G. 
BMMcL acknowledges support from the ITAMP visitor's program, 
Queen's University Belfast for the award 
of a Visiting Research Fellowship (VRF) and the hospitality 
of the University of Georgia during recent research visits.
ITAMP is supported in part by a grant from the NSF to the Smithsonian
Astrophysical Observatory and Harvard University. 
We would like to thank Dr. Charles W. Bauschlicher from NASA Ames, 
for sending us his data  in numerical format. 
Grants of computational time  at the National Energy Research  
Scientific Computing Center (NERSC) in Berkeley, 
CA, USA  and at the High Performance  Computing Center  Stuttgart (HLRS) of the 
University of Stuttgart, Stuttgart, Germany are gratefully acknowledged.
This research made use of the NASA Astrophysics Data System.



\bibliographystyle{mnras}

\bibliography{sio-ra} 






\bsp	
\label{lastpage}
\end{document}